%% file: flatftheory_sub2.tex
\documentclass[11pt]{article}
\pdfoutput=1

\usepackage{stolenstyle}

\usepackage{amsmath,epsf,amssymb,latexsym,amsthm,setspace,bbm,array,pifont,enumerate}

\input{basedefs2}

\theoremstyle{definition}

\usepackage{tikz}

\usetikzlibrary[shapes.geometric]
\usetikzlibrary{positioning} 
\usetikzlibrary{calc,intersections,through,backgrounds}
\usetikzlibrary{cd}

\tikzset{>=stealth}
\tikzset{every picture/.style={very thick}}

\def\Spind{{\Spin_{\text{d}}}}
\def\Spindp{{\Spin'_{\text{d}}}}

\def\bF{{{{\boldsymbol{F}}}}}

\def\sleft{\text{\tiny{L}}}
\def\sright{\text{\tiny{R}}}

\def\preprint{ ~~ }

\title{Flat F-theory and friends}
\author[a] {Peng Cheng,}
\author[b] {Ilarion V.~Melnikov,}
\author[c] {Ruben Minasian}
\affiliation[a] {Arnold Sommerfeld Center, LMU M\"unchen, Theresienstra{\ss}e 37, 80333 M\"unchen, Germany}
\affiliation[b] {Department of Physics and Astronomy,
James Madison University,
Harrisonburg, VA 22807, USA}
\affiliation[c] {Institut de Physique Th{\'e}orique,  
Universit{\'e} Paris-Saclay, CNRS, CEA, F-9119, Gif-sur-Yvette, France
}

\emailAdd{Peng.Cheng@physik.uni-muenchen.de}
\emailAdd{melnikix@jmu.edu}
\emailAdd{ruben.minasian@ipht.fr}

\abstract{We discuss F-theory backgrounds associated to flat torus bundles over Ricci-flat manifolds.  In this setting the F-theory background can be understood as a IIB orientifold with a large radius limit described by a supersymmetric compactification of IIB supergravity on a smooth, Ricci flat, but in general non-spin geometry.  When compactified on an additional circle these backgrounds are T-dual to IIA compactifications on smooth non-orientable manifolds with a $\text{Pin}^-$ structure.}

\begin{document}

\maketitle

\section{Introduction} \label{s:Introduction}

Given a type IIB string theory compactified on a manifold $Y\times\R^{1,d-1}$ with a discrete symmetry action $G$, we can orientifold the construction to obtain a compactification $Y\times\R^{1,d-1}/H$, where  $H = G' \cup G \Pi$, and $\Pi$ denotes a worldsheet parity operation~\cite{Angelantonj:2002ct,Gimon:1996rq,Dabholkar:1997zd}.  Typically the $G$-action will have a non-empty fixed locus on $Y$, and this locus will mark the locations of orientifold planes.  The orientifold planes source a tadpole that can be canceled by adding D-branes on top of the orientifold planes, and the resulting theory can preserve some reduced amount of supersymmetry.  As shown in the classic work~\cite{Sen:1996vd,Sen:1997gv}, such constructions can often be identified with certain limits of related F-theory compactifications.

The introduction of open string degrees of freedom is a central feature of this construction that leads to much of the beauty and intricate structure of the resulting theory.  In this note we point out an interesting corner of the orientifold landscape:  when $G$ acts freely on $Y$ no orientifold planes are required, so that the resulting theories can be understood without additional D-branes or open strings.  Furthermore, in the large radius limit the effective theory is described as compactification of IIB supergravity on  $X=Y/G$---a smooth space that need not admit a spin structure.  These compactifications also have an F-theory interpretation in terms of an elliptically fibered Calabi-Yau manifold $\pi : Z \to X$, where the fibration corresponds to a flat $\SL(2,\Z)$ bundle, so that the F-theory elliptic fiber is locally constant.

While well-known to experts since the inception of F-theory---see for example sections 6.3 and 6.4 of~\cite{Morrison:1996pp}, there are a number of reasons why these examples deserve more attention.  First, they provide very simple examples of supersymmetric string compactification on non-spin manifolds in IIB or non-orientable manifolds in IIA, where a crucial role is played by the lift of a bosonic duality group, such as $\SL(2,\Z)$ of the ten-dimensional IIB string theory, to the action on the full theory, including the spacetime fermion fields~\cite{Pantev:2016nze,Tachikawa:2018njr,Debray:2021vob}.  Second, they illustrate the importance of global topological features, such as certain non-trivial torsion classes in cohomology, and allow us to probe these features directly on the string worldsheet.  Third, they provide additional examples of perturbative string compactifications with reduced supersymmetry in various dimensions.  Finally, their simple structure makes these vacua ideal for detailed explorations of string duality.

In the rest of this note we first illustrate the idea with a well-known example based on~\cite{Ferrara:1995yx}.  We then describe a number of related examples in diverse dimensions, point out a relationship to compactification of the IIA string on non-orientable manifolds, and we conclude with a brief outlook.  There is a small geometric appendix.

\section*{Acknowledgements} PC is partially supported by the DFG Excellence Strategy EXC-2094 390783311.  IVM's work is supported in part by the Humboldt Research Award and the Jean d'Alembert Program at the University of Paris--Saclay, as well as the Educational Leave program at James Madison University.  RM is partially supported by ERC grants 772408-Stringlandscape and 787320-QBH Structure, as well as the Humboldt Foundation.    We thank  F.~Bonetti, I.~Brunner, M.~Del Zotto, A.~Font, J.~Gray, and I.~Saberi for useful discussions.  This work initiated during a stay at the Albert Einstein Institute (Max Planck Institute for Gravitational Physics), and we are grateful to the AEI for its generous hospitality.

\section{The Enriques orientifold}

\subsection{The Enriques geometry as a quotient}
The starting point for our discussion is compactification of IIB string theory on a smooth K3 surface $Y$~.  The massless spectrum is well-known~\cite{Romans:1986er,Hull:1994ys,Witten:1995ex}: it is (2,0) supergravity with $21$ chiral tensor multiplets and a moduli space $\cM_{5,21} \simeq \Gr(5,21)/\GO(\Gamma_{5,21})$, where  $\Gr(n_1,n_2)$ is the Grassmannian 
\begin{align}
\Gr(n_1,n_2) = \GO(n_1,n_2) /\GO(n_1)\times\GO(n_2)~,
\end{align}
and $\GO(\Gamma_{n_1,n_2})$ is the group of isometries of the integral lattice $\Gamma_{n_1,n_2}$ with signature $(n_1,n_2)$.

Working in the RNS formulation, the string-perturbative regime is described by a worldsheet theory with (4,4) superconformal symmetry and central charges $c_{\sleft} = c_{\sright} = 6$, and if we further take a large radius limit, then the worldsheet theory has a description as a non-linear sigma model with target space $Y$.  This SCFT has moduli space $\cM_{4,20}$ defined in an analogous fashion to $\cM_{5,21}$~\cite{Seiberg:1988pf,MR1416354}.  In the large radius limit we can decompose $\cM_{4,20}$ into the moduli space of Einstein volume $1$ metrics $g$ on K3---this is isomorphic to~$\cM_{3,19}$, a choice of closed B-field  $B \in H^2(Y,\R)/ H^2(Y,\Z)$, and the K3 volume $V$~\cite{Aspinwall:1996mn}.   $Y$ is a hyper-K\"ahler manifold with holonomy $\SU(2) = \Sp(1)$ and consequently has $3$ covariantly constant forms $j \in H^{1,1}(Y,\C)$, $\omega \in H^{2,0}(Y,\C)$, and $\omegab \in H^{0,2}(Y,\C)$ which obey
\begin{align}
j \wedge j & = \ff{1}{2} \omega \wedge \omegab = \dVol_g(Y)~, &
j\wedge\omega & = \omega\wedge \omega = \omegab\wedge\omegab = 0~.
\end{align}
It is useful to think of the holonomy group of $Y$ in terms of the reduction of the holonomy of $Y$ from the generic  $\SO(4) = \left\{ \SU(2)_1\times\SU(2)_2\right\}/\Z_2$.  Here the quotient is by the diagonal central element, and we can identify the holonomy action with $\SU(2)_2$.  The forms $j,\omega,\omegab$ are singlets with respect to $\SU(2)_2$ but transform in $\rep{3}$ of $\SU(2)_1$, which describes the hyper-K\"ahler rotations.  We can think of $Y$ as a Calabi-Yau manifold with trivial canonical bundle $K_Y$ and $h^1(\cO_Y) = 0$.

There is a locus in the metric moduli space $\cM_E \subset \cM_{3,19}$, where $Y$ admits a freely-acting K\"ahler isometry group $G_E \simeq \Z_2$~\cite{Barth:2004ne}, and its generator $U_E$ acts on the triplet of forms as
\begin{align}
U_E \cdot j & = j ~,&
U_E \cdot \omega & = - \omega~,&
U_E \cdot \omegab & = - \omegab~.
\end{align}
This free action is known as the Enriques involution, and the quotient $X = Y/G_E$ is an Enriques surface--- a smooth compact K\"ahler Ricci-flat surface with a torsion canonical bundle:  $K_X^{\otimes 2} \simeq \cO_X$, but $K_X \neq \cO_X$.  Recall that a complex manifold $X$ is spin if and only if  $c_1(K_X) = 0 \mod 2$~(see, e.g. \cite{Lawson:1989sp}); thus, the Enriques surface $X$ is not spin.  We can see this explicitly in the quotient construction, as the action of $U_E$ on the hyper-K\"ahler forms and on the covariantly constant spinors is by the element $i\sigma_3 \in \SU(2)_1$.  Letting $J^3$ and $J^\pm$ be the generators of $\SU(2)_1$ with the non-zero commutators
\begin{align}
\CO{J^3}{J^\pm} & = \pm 2 J^\pm~,&
\CO{J^\pm}{J^\mp} & = \pm J^3~,
\end{align}
the action can also be expressed as $e^{i\ff{\pi}{2} J^3}$.   It is no coincidence that the quotient also enlarges the holonomy group: we have $\pi_1(X) = \Z_2$, and the quotient metric $g_X$ has holonomy group $\text{Hol}(X,g_X) = \SU(2) \rtimes \Z_2$.  The first factor is the null-homotopic holonomy inherited from $Y$, while the second factor is associated to the holonomy around a loop $\gamma$ that generates $\pi_1(X)\simeq \Z_2$.\footnote{The semi-direct product in the holonomy group is discussed in the appendix.}

The last geometric point we will need is the $G_E$ action on the lattice $H^2(Y,\Z)  = \Gamma_{3,19}$ with respect to the Enriques action~\cite{Barth:2004ne,Ferrara:1995yx}.  Decomposing
\begin{align}
\Gamma_{3,19} = (\Gamma_{0,8} + \Gamma_{1,1}) + (\Gamma_{0,8}+\Gamma_{1,1}) +\Gamma_{1,1}~,
\end{align}
the generator acts as 
\begin{align}
U_E & = \begin{pmatrix} 0 & \iden_{10} & 0 \\ \iden_{10} & 0 & 0 \\ 0 & 0 & -\iden_{2} \end{pmatrix}~,
\end{align}
and consequently the spaces of dual and anti-self-dual forms decompose into even and odd components as
\begin{align}
\label{eq:evenoddcohomology}
H^2_+(Y,\R) & = \R_{\text{e}} \oplus \left( \R^2\right)_{\text{o}}~,&
H^2_-(Y,\R)  & =\left(\R^{9}\right)_{\text{e}} \oplus  \left( \R^{10} \right)_{\text{o}}~.
\end{align}

\subsection{The Enriques symmetry in the SCFT}
Denote by $\cS[Y]$ the superconformal theory associated to the non-linear sigma model on $Y$ equipped with Einstein metric $G = \sqrt{V} g$, where $g$ is an Einstein metric with volume $1$, and a closed $B$-field $B$.  Choosing $g \in \cM_E$ and setting $B = 0$, we expect $\cS[Y]$ to admit a global $\Z_2$ symmetry, which we will also call $G_E$.\footnote{It is not necessary to set $B=0$.  The most general choice is to choose $B$ so that the difference $g_E\cdot B  - B$ is contained in the image of $H^2(Y,\Z)$ inside $H^2(Y,\R)$.}  Its generator is a unitary operator which we will also denote $U_E$.  Although we will use some features of the large radius limit, we will phrase our discussion as much as possible in terms of the SCFT properties to emphasize that in principle we have a construction that works at generic points on the Enriques locus.  This includes solvable points, such as $T^4/\Z_n$ orbifolds and Gepner points.

In the NS-NS sector the action of the generator $U_E$ is a combination of two operations.  The first, which we call $x_E$, is a $\Z_2$ action on the (4,4) primary fields of the SCFT, while the second is a rotation generated by the global charges of the $\widehat{\su}(2)_{\sleft} \oplus \widehat{\su}(2)_{\sright}$ Kac-Moody algebra contained in the superconformal algebra:
\begin{align}
 \exp\left[ \ff{i\pi}{2} J^3_{\sleft 0} + \ff{i\pi}{2} J^3_{\sright 0} \right]~.
\end{align}
The symmetry generator is then
\begin{align}
U_E & = x_E \exp\left[ \ff{i\pi}{2} J^3_{\sleft 0} + \ff{i\pi}{2} J^3_{\sright 0} \right]~.
\end{align}
It is not hard to see that $U_E^2 = \iden$ in the NS-NS sector.  The only subtlety is in the factor $\exp\left[ \ff{i\pi}{2} J^3_{\sleft 0} + \ff{i\pi}{2} J^3_{\sright 0} \right]$.   To see that it squares to identity note that the primary fields in the NS-NS sector transform either in $(\rep{1},\rep{1})$ or $(\rep{2},\rep{2})$ of $\widehat{\su}(2)_{\sleft} \oplus \widehat{\su}(2)_{\sright}$.\footnote{ At $c=6$ unitarity only allows the representations $\rep{1}$ and $\rep{2}$ to appear, and by considering a limiting case such as $Y = T^4/
\Z_2$, where the theory is solvable, we see that in fact the only representations possible are $(\rep{1},\rep{1})$ or $(\rep{2},\rep{2})$.   Unitarity constraints on representations of the N=4 superconformal algebra are explored in detail in~\cite{Eguchi:1988af}.}  In the latter case the action on the (4,4) primary field is determined by specifying the action of $x_E$ on, say, the state of charge $+1$ with respect to $J^{3}_{\sleft 0}$ and $-1$ with respect to $J^{3}_{\sright 0}$.

For example we can consider the operators that contribute to the massless spacetime spectrum.  In addition to the identity operator these are (4,4) primary operators with $h_{\sleft} = h_{\sright} = 1/2$, transforming in $(\rep{2},\rep{2})$ of $\SU(2)_{\sleft}\times\SU(2)_{\sright}$.  Picking a subgroup $\GU(1)_{\sleft} \times\GU(1)_{\sright} \subset \SU(2)_{\sleft}\times\SU(2)_{\sright}$ with corresponding Kac-Moody currents $J^3_{\sleft,\sright}$ as above, we find that these operators transform in a quartet consisting of $N=2$ (anti)-chiral primary operators on the left and the right:
\begin{equation}
\begin{tikzcd}
\cO^{\text{ca}}\arrow[r, "J^{+}_{\sright}"]  \arrow[d,"J^{-}_{\sleft}"]& 
\cO^{\text{cc}} \arrow[d,"J^{-}_{\sleft}"]~ \\
\cO^{\text{aa}} \arrow[r,"J^{+}_{\sright}"] & 
\cO^{\text{ac}}
\end{tikzcd}
\end{equation}
Here $\cO^{\text{cc}}$ is an operator that is chiral-primary both on the left and right.  We can be very explicit when the SCFT is a non-linear sigma model in a large radius limit, i.e. $g$ is at a smooth point in $\cM_E$ and $\sqrt{V} \gg \alpha'$.  In this case taking complex coordinates for the bosonic target space fields $Z^i$, $i=1,2$, with their left- and right-moving superpartners Weyl fermions $\Psi^i_{\sleft}$ and $\Psi^i_{\sright}$, as well as their target-space complex conjugates, we find that there are two types quartets contributing to the massless spectrum:
\begin{enumerate}
\item the \textit{self-dual quartet} with components
\begin{equation}
\begin{tikzcd}
j_{i\jb} \Psi^{i}_{\sleft} \Psib^{\jb}_{\sright} \arrow[r, "J^{++}_{\sright}"]  \arrow[d,"J^{--}_{\sleft}"]& 
-\omega_{ij} \Psi^{i}_{\sleft} \Psi^j_{\sright} \arrow[d,"J^{--}_{\sleft}"]~ \\
-\omegab_{\ib\jb} \Psib^{\ib}_{\sleft}\Psib^{\jb}_{\sright} \arrow[r,"J^{++}_{\sright}"] & 
-j_{j\ib} \Psib^{\ib}_{\sleft} \Psi^j_{\sright}
\end{tikzcd}~,
\end{equation}
\item the $19$ \textit{anti-self-dual quartets}, each with components
\begin{equation}
\begin{tikzcd}
\cA^{\text{ca}}_{i\jb} \Psi^{i}_{\sleft} \Psib^{\jb}_{\sright} \arrow[r, "J^{++}_{\sright}"]  \arrow[d,"J^{--}_{\sleft}"]& 
\cA^{\text{cc}}_{ij} \Psi^{i}_{\sleft} \Psi^j_{\sright} \arrow[d,"J^{--}_{\sleft}"]~ \\
\cA^{\text{aa}}_{\ib\jb} \Psib^{\ib}_{\sleft}\Psib^{\jb}_{\sright} \arrow[r,"J^{++}_{\sright}"] & 
\cA^{\text{ac}}_{\ib j}  \Psib^{\ib}_{\sleft} \Psi^j_{\sright}
\end{tikzcd}~,
\end{equation}
where $\cA^{\text{ca}} = \cA^{\text{ca}}_{i\jb} dZ^i\wedge d\Zb^{\jb}$ is a harmonic anti-self-dual form\footnote{Anti-self-duality of $\cA^{\text{ca}}$ is equivalent to $j \wedge \cA^{\text{ca}} = 0$.}, and
\begin{align}
\cA^{\text{cc}}_{ij} & = -\cA^{\text{ac}}_{i\kb} g^{\kb k} \omega_{kj}~,&
\cA^{\text{aa}}_{\ib\jb} & = -\omegab_{\ib \kb} g^{\kb k} \cA^{\text{ac}}_{k\jb}~,&
\cA^{\text{ac}}_{\ib j} & = \omegab_{\ib \mb} g^{\mb m} \cA_{m\kb} g^{\kb k} \omega_{kj}~.
\end{align}
It is not hard to show that anti-self-duality of $\cA^{\text{ca}}$ is equivalent to  $\cA^{\text{cc}}_{ij} = \cA^{\text{cc}}_{ji}$.
\end{enumerate}
Since we know the action of $x_E$ on the cohomology of $Y$, we can now decompose these quartets according to the (2,2) superconformal algebra that commutes with the action of $G_E$.  The self-dual quartet decomposes as
\begin{align}
\text{SD quartet} = (\text{c},\text{c})_{\text{o}} \oplus (\text{a},\text{a})_{\text{o}} \oplus (\text{c},\text{a})_{\text{e}} \oplus (\text{a},\text{c})_{\text{e}}~.
\end{align}
An anti-self-dual quartet has different decompositions, depending on whether $x_E \cdot \cA^{\text{ca}} = \pm \cA^{\text{ca}}$:
\begin{align}
\text{ASD quartet}^+ = (\text{c},\text{c})_{\text{o}} \oplus (\text{a},\text{a})_{\text{o}} \oplus (\text{c},\text{a})_{\text{e}} \oplus (\text{a},\text{c})_{\text{e}}~, \nonumber\\
\text{ASD quartet}^- = (\text{c},\text{c})_{\text{e}} \oplus (\text{a},\text{a})_{\text{e}} \oplus (\text{c},\text{a})_{\text{o}} \oplus (\text{a},\text{c})_{\text{o}}~. \nonumber\\
\end{align}

With this set up we can define an orbifold $\cS[Y] /G_E$.  The resulting theory will only have (2,2) supersymmetry because the currents $J^\pm_{\sleft}$ and $J^\pm_{\sright}$ will be projected out.  It will have non-trivial (c,c) and (a,c) rings (and their conjugates), each containing the identity operator, $10$ operators with dimension $h_{\sleft}=h_{\sright} = 1/2$, and one operator of top dimension $h_{\sleft} = h_{\sright} = 1$ represented by $J_{\sleft}^+ J_{\sright}^+$ in the (c,c) ring and $J_{\sleft}^- J_{\sright}^+$ in the (a,c) ring.  Crucially for our purposes the projection will also remove the spectral flow operators that we would usually use to construct the chiral Ramond sectors and define a chiral GSO projection.  This is the worldsheet analogue of the spacetime statement that the Enriques orbifold projects out the covariantly constant spinors.

\subsection{Some RNS details}
In order to obtain supersymmetric vacua based on the Enriques orbifold we will need to enlarge the action of $G_E$ to act on the entire world-sheet theory, including the Minkowski degrees of freedom and the superconformal ghosts.  To describe the structure we will use the covariant formulation of the RNS string, and we will not need too many details beyond the usual textbook treatment~\cite{Polchinski:1998rr}.

The first additional ingredient we will need is a description of the Ramond sectors associated to the $\cS[Y]$ SCFT.  These are easily constructed by spectral flow once we pick a (2,2) subalgebra of the (4,4) algebra.  In our case there is a preferred choice of a (2,2) sub-algebra --- the sub-algebra that commutes with the action of $U_E$.  Then, by the usual construction~\cite{Lerche:1989uy}, we obtain an isomorphism between the Ramond ground states of the theory and the chiral primary operators in the NS sector.  Since the theory has chiral spectral flow, we can do this separately on the left and right.  For example, left spectral flow of the identity operator and $J^+_{\sleft}$ leads to spin fields $\Sigma^\pm_{\sleft}$ which have the OPE
\begin{align}
\Sigma_{\sleft}^{A} (z_1) \Sigma_{\sleft}^{B}(z_2) & \sim \frac{\ep^{AB}}{\sqrt{z_{12}}} + \frac{1}{2} \sqrt{z_{12}} \delta^{A,B} J_\sleft^{B}~.
\end{align}
Moreover these are primary with respect to $\widehat{\su}(2)_{\sleft}$, transforming in a doublet.  These universal spin fields will play a role in constructing spacetime supercharges and a universal tensor multiplet in spacetime.  On the other hand, chiral primary fields with $h_{\sleft} = 1$ lead to left-moving Ramond ground states that are singlets with respect to $\widehat{\su}(2)_{\sleft}$.  More generally, every NS-NS field has an image in the R-NS sector determined by the left-moving spectral flow, and we have the same construction on the right.  Finally, the internal theory has chiral fermion number operators that are represented by
\begin{align}
e^{i \pi F_{\sleft}^{\text{int}}} & = e^{i \pi J^3_{\sleft 0}}~,&
e^{i \pi F_{\sright}^{\text{int}}} & = e^{i \pi J^3_{\sright 0}}~.
\end{align}

Next we describe the Minkowski degrees of freedom: these are bosons $X^\mu$ for $\R^{1,5}$, and their Majorana-Weyl superpartners $\cX^\mu_{\sleft}$, $\cX^\mu_{\sright}$.   The currents $\Lambda^{\mu\nu} = : \cX_L^\mu \cX_L^\nu$ generate a Lorentz current algebra $\widehat{\so}(1,5)_1$, and we can construct spin fields $\cS_{\sleft a}$ that transform in the reducible $\rep{8}$ representation of the associated $\Spin(1,5)$.  Conformal invariance, together with the spacetime Lorentz symmetry constrain the OPEs of the spin fields:
\begin{align}
\cX^\mu(z_1) \cS_a(z_2) &\sim \frac{1}{\sqrt{2 z_{12}}} \Gamma^\mu_{ab} \cS_b(z_2) ~,&
\cS_a(z_1) \cS_b(z_2) &\sim\frac{\cC_{ab}}{z_{12}^{3/4}} + \frac{ (\cC\Gamma^\mu)_{ab}}{\sqrt{2} z_{12}^{1/4}} \cX_{\mu}(z_2)~.
\end{align}
Here the $(\Gamma^\mu)_{ab}$ are the generators of the non-trivial irreducible representation of $\text{Cl}(1,5)$, and $\cC$ is the symmetric charge conjugation matrix that anticommutes with the chirality matrix $\Gamma_6 = -\Gamma^0 \Gamma^1\cdots\Gamma^5$.  The action of the fermion number operator $e^{i\pi F^{\text{min}}_{\sleft}}$ on the spin fields can be determined in a number of ways, for example by bosonizing the fermions $\cX^\mu_{\sleft}$.  The result is
\begin{align}
e^{i\pi F^{\text{min}}_\sleft} \cS_a(z) e^{-i\pi F^{\text{min}}_\sleft} =(e^{-i\pi \Gamma_6/2} )_{ab} \cS_b(z) ~.
\end{align}
In other words, the spin fields $\cS^\pm$ satisfying $\Gamma_6 \cS^\pm = \pm \cS^\pm$ carry fermion number $\mp i$.  It will be convenient for us to decompose the reducible $\rep{8}$ representation of $\spin(1,5)$ into its irreducible Weyl components.  In this representation
\begin{align}
\Gamma_6 & = \begin{pmatrix} \iden_4 &  0 \\ 0 &\-\iden_4 \end{pmatrix}~,&
\cC & = \begin{pmatrix} 0 & \iden_{4} \\ \iden_{4} & 0 \end{pmatrix}~, &
\cC\Gamma^\mu & = \begin{pmatrix} \sigma^\mu &  0 \\ 0 &\sigmat^\mu \end{pmatrix}~,
\end{align}
with $\sigma$ and $\sigmat$ anti-symmetric matrices.

Finally, we need a little bit of information on the ghost sector. The relevant ghost sector operators are constructed from the fields that bosonize the (1,1) $\beta_{{\sleft}}\gamma_{{\sleft}}$ and $\beta_{\sright} \gamma_{\sright}$ currents as $\p \varphi_{\sleft}$,  $\pb \varphi_{\sright}$.  We will need the operators $e^{-s \varphi_{\sleft}}(z)$ and $e^{-s \varphi_{\sright}}(\zb)$, which have conformal dimensions $(s-s^2/2,0)$ and $(0,s-s^2/2)$ and fermion numbers
\begin{align}
e^{i\pi F^{\text{gh}}_{\sleft}} e^{-s\varphi_{\sleft}} e^{-i\pi F^{\text{gh}}_{\sleft}} &= e^{-i\pi s}e^{-s\varphi_{\sleft}} ~,&
e^{i\pi F^{\text{gh}}_{\sright}} e^{-s\varphi_{\sright}} e^{-i\pi F^{\text{gh}}_{\sright}} &= e^{-i\pi s}e^{-s\varphi_{\sright}} ~.
\end{align}

With these ingredients in hand, we can construct the R-NS, NS-R, and R-R sectors and impose the IIB GSO projection
\begin{align}
e^{i\pi F^{\text{tot}}_{\sleft} }& = 1~,&e^{i\pi F^{\text{tot}}_{\sright} }& = 1~.
\end{align}
In what follows we will focus on two aspects of the full SCFT with GSO projection:  the massless spectrum and the worldsheet realization of the spacetime supersymmetry algebra.  We begin our discussion with the latter, following the classic treatment~\cite{Friedan:1985ge,Banks:1988yz}:  by stripping the gravitino vertex operators in the (-1/2,0) and (0,-1/2) pictures we obtain chiral worldsheet currents
\begin{align}
\cJ^{A}_{\sleft \alpha} &= e^{-\varphi_{\sleft}/2} \Sigma^A_{\sleft} \cS^+_{\sleft \alpha}~,&
\cJ^{A}_{\sright \alpha} &= e^{-\varphi_{\sright}/2} \Sigma^{A}_{\sright} \cS^+_{\sright \alpha}~.
\end{align}
Using the preceding analysis we calculate
\begin{align}
\cJ^A_{\sleft\alpha} (z_1) \cJ^B_{\sleft \beta} (z_2) \sim \frac{1}{z_{12}} \ep^{AB} \sigma^\mu_{\beta\alpha} e^{-\varphi_{\sleft}} \cX_{\sleft\mu}(z_2)~,
\end{align}
where we recognize $\cP_\mu(z_2) = e^{-\varphi_{\sleft}} \cX_\mu(z_2)$ as the conserved current associated to spacetime translations in the (-1,0) picture.  This means the conserved charges $\cQ^{A}_{\sleft \alpha} = \oint \ff{dz}{2\pi i} \cJ^A_{\sleft \alpha}$ and $P_{\sleft\mu} =  \oint \ff{dz}{2\pi i} \cP_{\sleft\mu}$ and their right-moving counter-parts obey the supersymmetry algebra\footnote{Notice that the odd pole in the $\cJ_L$--$\cJ_L$ OPE, together with anti-symmetry of $\sigma$ implies that the $\cQ^A_{\sleft\alpha}$ are anticommuting; moreover the operators $P_{\sleft\mu}$ and $P_{\sright\mu}$ both represent the spacetime translation current, expressed in different pictures.  }
\begin{align}
\AC{ \cQ^A_{\sleft \alpha}}{\cQ^B_{\sleft\beta}} &= \ep^{AB}\sigma^\mu_{\beta\alpha} P_{\sleft\mu}~,&
\AC{ \cQ^A_{\sright \alpha}}{\cQ^B_{\sright\beta}} &= \ep^{AB}\sigma^\mu_{\beta\alpha} P_{\sright\mu}.
\end{align}
This is exactly the supersymmetry algebra of (2,0) supergravity in six dimensions.

Having introduced the Ramond sectors, we also note that the full worldsheet theory enjoys additional symmetries generated by the spacetime fermion number operators $e^{i\pi \bF_{\sleft}}$ and $e^{i\pi\bF_{\sright}}$.  The latter assigns charge $-1$ to NS-R and R-R fields and leaves the other sectors invariant, and the former has an analogous action on the R-NS and R-R fields.  We obtain the total six-dimensional spacetime fermion number by combining these, and it is then not hard to verify
\begin{align}
\label{eq:totalspacetimefermion}
e^{i\pi \bF} = e^{i\pi \bF_{\sleft}} e^{i\pi \bF_{\sright}}= e^{i \pi J^3_{\sleft0} + i\pi J^3_{\sright 0}}~.
\end{align}

\subsection{Worldsheet parity and a supersymmetric Enriques action}
The Enriques action in the NS-NS sector has a lift to the R-NS and NS-R sectors, and because the internal spin fields transform in doublets of the $\SU(2)_{\sleft}\times\SU(2)_{\sright}$ it is easy to see that the spacetime supercharges are charged with respect to $U_E$, and, moreover, $U_E^2 \neq \iden$ on these sectors. However, there is a simple modification of the quotient that will preserve spacetime supersymmetry through an orientifold construction (see ~\cite{Angelantonj:2002ct,Dabholkar:1997zd} for thorough reviews).  

We can choose moduli of the internal theory, for example by setting $B=0$, so that in addition to the $U_E$ action described above, the IIB worldsheet theory has a parity symmetry which acts on the worldsheet coordinates as $z\mapsto \zb$, and there is a unitary Hermitian operator $\Pi$ that takes each Virasoro primary field to a Virasoro primary field of the same weight and opposite spin:  $\cO(z,\zb) \mapsto \Pi \cO(z,\zb) \Pi = \overline{\cO}(\zb,z)$, and  the action preserves the OPE and correlation functions.  The modified Enriques action is then taken to be
\begin{align}
\Ut_E =e^{i\pi\bF_{\sleft}}  U_E \Pi~.
\end{align}
Using~(\ref{eq:totalspacetimefermion}) it is then easy to see that $\Ut_E^2 = \iden$ on all sectors.  Furthermore $\Ut_E$ now preserves a number of fundamental structures including a diagonal Virasoro algebra with $c=12$ generated by the modes $\Lh_m = L_{\sleft m}+ L_{\sright m}$, and a level $2$ ``even'' Kac-Moody algebra $\widehat{\su}(2)_2$ generated by the modes
\begin{align}
\Jh_m^3 & = J^3_{\sleft} + J^3_{\sright}~, &
\Jh^\pm_m  & = J^\pm_{\sleft m} - J^\pm_{\sright m}~.
\end{align}
The spacetime supersymmetry currents now transform as
\begin{align}
\Ut_E \cJ^\pm_{\sleft}(z) \Ut_E &= \pm i \cJ^\pm_{\sright\alpha}(z)~, &
\Ut_E \cJ^\pm_{\sright}(\zb) \Ut_E &= \mp i \cJ^\pm_{\sleft\alpha}(\zb)~,
\end{align}
so that if we define the linear combinations of the charges
\begin{align}
\cQ^{+}_{\text{e}} & = \cQ^+_{\sleft} + i \cQ^+_{\sright}~, & \cQ^+_{\text{o}} & = \cQ^+_{\sleft} - i \cQ^+_{\sright}~, \nonumber\\
\cQ^{-}_{\text{e}} & = \cQ^-_{\sleft} - i \cQ^-_{\sright}~,& \cQ^-_{\text{o}} & = \cQ^-_{\sleft} + i \cQ^-_{\sright}~,
\end{align}
then
\begin{align}
\Ut_E \cQ^A_{\text{e}} \Ut_E &=\cQ^A_{\text{e}}~,&
\Ut_E \cQ^A_{\text{o}} \Ut_E &=-\cQ^A_{\text{o}}~,
\end{align}
and each pair transforms as a doublet with respect to the $\SU(2)_{\text{e}}$ symmetry generated by the ``even'' Kac-Moody algebra.

Using the set up of the previous section it is now easy to characterize the massless spacetime spectrum of IIB on K3 in terms of the $\Ut_E$ action.    We can summarize the results in terms of (c,a) fields of the internal theory and the resulting spacetime multiplets:
\begin{enumerate}
\item vacuum:
\begin{align}
\{\text{(2,0) gravity}\} &= \{\text{(1,0) gravity}\}_{\text{e}} \oplus \{\text{(1,0) gravitinos}\}_{\text{o}} \nonumber\\
\{\text{(2,0) tensor}\} &= \{\text{(1,0) tensor}\}_{\text{o}} \oplus \{\text{(1,0) hyper}\}_{\text{e}}~.
\end{align}
\item $\cO^{\text{ca}}[g]$:
\begin{align}
\{\text{(2,0) tensor}\} &= \{\text{(1,0) tensor}\}_{\text{o}} \oplus \{\text{(1,0) hyper}\}_{\text{e}}~.
\end{align}
\item $\cO^{\text{ca}}[\cA]$, $\cA$ anti-self-dual and $g_E$-even ($9$ of these):
\begin{align}
\{\text{(2,0) tensor}\} &= \{\text{(1,0) tensor}\}_{\text{e}} \oplus \{\text{(1,0) hyper}\}_{\text{o}}~.
\end{align}
\item $\cO^{\text{ca}}[\cA]$, $\cA$ anti-self-dual and $g_E$-odd ($10$ of these):
\begin{align}
\{\text{(2,0) tensor}\} &= \{\text{(1,0) tensor}\}_{\text{o}} \oplus \{\text{(1,0) hyper}\}_{\text{e}}~.
\end{align}
\end{enumerate}

\subsection{An orientifold without open strings and its supergravity limit}
We now use the $\Ut_E$ action on the IIB string worldsheet to define a $\Z_2$ orientifold.  The $\Ut_E$--even modes include the spacetime supercharges $\cQ_{\text{e}}^A$ transforming in a doublet of $\SU(2)_{\text{e}}$, and taking the projection in the untwisted sector leads to a massless spectrum consisting of (1,0) supergravity coupled to $9$ tensor multiplets and $12$ hypermultiplets---exactly the massless spectrum of F-theory compactification on $Z = (Y\times T^2)/\Z_2$~\cite{Morrison:1996pp,Ferrara:1996wv}---the Enriques Calabi-Yau $3$-fold familiar from the Borcea-Voisin construction of mirror pairs~\cite{Voisin:1993mir,Borcea:1996mxz} and introduced in the context of string duality in~\cite{Ferrara:1995yx}.  The same spectrum arises in the closed string sector of the orientifold introduced in~\cite{Dabholkar:1996zi} (there it is also accompanied by open string modes and a non-trivial gauge symmetry of rank at least $8$), and the same spectrum is reproduced in the $\Z_4$ orientifold of $T^4$ studied, among other models, in~\cite{Gimon:1996ay}.  Because our theory has neither orientifold planes or ordinary orbifold singularities, by contrast with both of these constructions, the orientifold we discuss is a consistent closed unoriented string (as is the model in~\cite{Gimon:1996ay}), but it is also entirely smooth, and thus has a supergravity description as a supersymmetric compactification of the IIB string on the Enriques surface $X$, a non-spin manifold.

To understand how it is possible to have a supersymmetric supergravity compactification on $X$, we recall that the fermions in IIB supergravity are not really spinors due to the non-trivial action of the duality group~\cite{Pantev:2016nze}.   Though it is often left unstated, this point is at the heart of orientifold and F-theory constructions.  The former have been discussed in great generality in~\cite{Distler:2009ri,Distler:2010an}.  Our worldsheet construction is an important special case of that discussion, in that the background preserves supersymmetry and also allows a complementary spacetime point of view, which we develop next.

The duality group of IIB has been recently discussed in~\cite{Tachikawa:2018njr,Debray:2021vob,Debray:2023yrs}, and the basic point relevant to us is that the bosonic duality symmetry $\SL(2,\Z)$ is enlarged to $\text{Pin}^+\GL(2,\Z)$ when the action on the fermions is taken into account.  Considering the Lorentz symmetry $\SO(1,9)$ together with the duality group, we find the corresponding faithful action on the fermions by a group we term $\Spind(1,9)$, which fits into the exact sequence
\begin{equation}
\begin{tikzcd}
1\ar[r] & \Z_2 \ar[r,"f"] & \Spind(1,9) \ar[r] & \SO(1,9) \times  \GL(2,\Z) \ar[r] & 1
\end{tikzcd}~.
\end{equation}
The duality group $\text{Pin}^+\GL(2,\Z)$ includes the spacetime fermion number $e^{i\pi \bF}$, and the map $f$ sends the $\Z_2$ generator to $(-\iden_{16}, e^{i\pi \bF}) \in \Spin(1,9) \times \text{Pin}^+\GL(2,\Z)$.  The possibilities for IIB compactification are then clearly enlarged if we allow to patch fields by $\Spind(1,9)$ rather than $\Spin(1,9)$ transition functions.  

A lift of a generic $\SL(2,\Z)$ element leads to transformations of the fermions by factors that depend on the axio-dilaton profile.  However, if we focus on perturbative backgrounds where the axio-dilaton remains constant and unconstrained, these factors drop out of the transformations, and we can restrict attention to the subgroup of perturbative duality symmetries $\cG_{\text{per}} \subset \text{Pin}^+\GL(2,\Z)$ generated by two generators $r = e^{i\pi \bF_{\sleft}}\Pi $ and $s = \Pi$~\cite{Dabholkar:1997zd,Debray:2021vob}:
\begin{align}
\cG_{\text{per}} \simeq D_{8} = \langle r,s ~|~ r^4 = s^2 = (rs)^2 = 1\rangle~.
\end{align}
The analogue of the bosonic duality group $\SL(2,\Z)$ is now $\Z_2\times\Z_2$ generated by $r$ and $s$ action on the bosonic fields, while $\cG_{\text{per}}$ is the lift that also includes the action on the fermions.  We then have
\begin{equation}
\label{eq:Spindp}
\begin{tikzcd}
1\ar[r] & \Z_2 \ar[r,"f"] & \Spindp(1,9) \ar[r] & \SO(1,9) \times  \Z_2\times\Z_2 \ar[r] & 1
\end{tikzcd}~.
\end{equation}

Working on a general ten-dimensional oriented spacetime $X$ then requires that the fermions transform in sections of a vector bundle $\cF\otimes T_X$, where the transition functions for $\cF$ are valued in $\Spindp(1,9)$.  If we take a good cover $\{\cU_\alpha\}_{\alpha \in I}$ for $X$, then over every patch we have
\begin{align}
\left.\cF\right|_{\cU_\alpha} = \cU_\alpha \times \R^{16} \otimes \R^2~.
\end{align}
A section $S$ of $\cF$ is then locally represented by $S_\alpha$, a real $16\times 2$ matrix, and on $\cU_\alpha\cap \cU_\beta \neq \emptyset$ the section patches by
\begin{align}
S_\alpha & = M_{\alpha\beta} S_\beta A_{\beta\alpha}~,
\end{align}
where the transition functions take value in $\Spindp(1,9)$:
\begin{align}
(M_{\alpha\beta}, A_{\beta\alpha}) \in \Spindp(1,9)~.
\end{align}
If we do not consider the non-trivial equivalence relation, we simply think of $M_{\alpha\beta} \in \Spin(1,9)$ and of $A_{\beta\alpha} \in \cG_{\text{p}}$.  The transition functions must satisfy the usual condition on a triple-overlap $\cU_\alpha\cap \cU_\beta\cap\cU_\gamma \neq \emptyset$
\begin{align}
(M_{\alpha\beta} M_{\beta\gamma} M_{\gamma\alpha}, A_{\alpha\gamma} A_{\gamma\beta} A_{\beta\alpha})  = \text{id} \in \Spindp(1,9)~.
\end{align}
A simple way to satisfy this is to have each factor be identity separately, so that $\cF = \cS_X \otimes \cR$, where $\cS_X$ is a spin bundle $\cS_X$ over $X$, while $\cR$ is a rank $2$ vector bundle over $X$ with structure group $\cG_{\text{p}}$.  The equivalence relation $(M,A)\sim(-M,-A)$ encoded in~(\ref{eq:Spindp}) allows for a more general possibility:
\begin{align}
M_{\alpha\beta} M_{\beta\gamma} M_{\gamma\alpha} &= \pm\iden_{16}~, &
A_{\alpha\gamma} A_{\gamma\beta} A_{\beta\alpha} &  = \pm \iden_{2}~.
\end{align}
The lower sign allows for $X$ to be non-spin.

Suppose now that the spacetime is of the form $\R^{1,5}\times Y$, where $Y$ is a four-dimensional compact oriented manifold.  In this case we obtain a decomposition
\begin{align}
\cF & = \cF_{+} \oplus \cF_{-}~,
\end{align}
corresponding to the decomposition
\begin{align}
\spin(1,9) &\supset \spin(1,5) \oplus\su(2)\oplus\su(2)~, \nonumber\\
\rep{16}   & = (\rep{4},\rep{1},\rep{2})\oplus(\rep{4}',\rep{2},\rep{1})~.
\end{align}
A  section  $\cS$ of $\cF_{+}$ has components $\cS_{aI}^{A}$, with $a = 1,\ldots, 8$, $I = 1,2$ denoting the $\spin(1,5)$ and $\su(2)$ indices, while $A=1,2$ is the $\cG_{\text{p}}$ index.  Reality of the underlying 10-dimensional field then implies
\begin{align}
\cS_{aI}^A = \cC_{ab} \ep_{IJ} \left(\cS_{bJ}^A\right)^\ast~,
\end{align}
where $\cC$ is the $6$-dimensional charge conjugation matrix, and $\ep_{IJ}$ is the invariant tensor of $\SU(2)$.

Since the transition functions are trivial in the $\R^{1,5}$ directions, the existence of fermions in $\cF_{+}$ boils down to the existence of a chiral $\Spindp(4)$ structure on $Y$.  Abusing notation slightly, we denote the corresponding bundle again by $\cF_+$, now with section locally having the form $\cS^A_I$, a $2\times 2$ complex-valued matrix.  If $Y$ is taken to be a K3 manifold, then clearly such a structure exists when the holonomy $\SU(2)$ is taken to be valued in the second factor, and we can take $\cS$ to be a constant $2\times 2$ matrix.

If we tune the moduli of $Y$ to the Enriques locus, then there is a K\"ahler isometry $\phi : Y\to Y$ such that $\phi$ is fixed-point free and $\phi \circ \phi = \text{id}$, and its action on tensors is by pull-back:  $U_E \cdot T = \phi^\ast T$.  The action on the frame bundle with local orthonormal basis $\boldsymbol{e}_a$ is simply $U_E \cdot \boldsymbol{e}_a(x) = M_a^b(x) \boldsymbol{e}_b(x)$, where $M_a^b \in \SO(4)$, and in terms of a holomorphic basis
\begin{align}
\label{eq:holoframe}
\boldsymbol{p}_1 & = \ff{1}{\sqrt{2}} \left( \boldsymbol{e_1} + i \boldsymbol{e}_4 \right)~,&
\boldsymbol{p}_2 & = \ff{1}{\sqrt{2}} \left( \boldsymbol{e_2} + i \boldsymbol{e}_3 \right)~,&
\boldsymbol{P} = \begin{pmatrix} \boldsymbol{p}_1 & \boldsymbol{p}_2 \\  -\overline{\boldsymbol{p}}_2& \overline{\boldsymbol{p}}_1 \end{pmatrix}~,
\end{align}
the $\SO(4)$ action by elements $g  = (M_1,M_2) \in \{ \SU(2)\times\SU(2)\}/\Z_2$ is
\begin{align}
g\cdot \boldsymbol{P} = M_1 \boldsymbol{P} M_2~.
\end{align}
It is easy to see that elements of the form $(\iden,M_2)$ preserve the hyper-K\"ahler forms
\begin{align}
j & = i \boldsymbol{p}_1 \wedge \overline{\boldsymbol{p}}_1 + i \boldsymbol{p}_2 \wedge \overline{\boldsymbol{p}}_2~,&
\omega & = 2 \boldsymbol{p}_1 \wedge\boldsymbol{p}_2~, &
\omegab & = 2 \overline{\boldsymbol{p}}_1 \wedge \overline{\boldsymbol{p}}_2~,
\end{align}
while those of the form $(M_1,\iden)$ rotate them as a triplet.  The Enriques action on the frame and the covariantly constant spinors is then by the element $U_E = (i\sigma_3,i\sigma_3)$,\footnote{ Here $\sigma_3$ is the usual Pauli matrix $\sigma_3 = \begin{pmatrix} 1 & 0 \\ 0 & -1\end{pmatrix}$.}:
\begin{align}
U_E \cdot \boldsymbol{P} &= \begin{pmatrix} -\boldsymbol{p}_1 & \boldsymbol{p}_2 \\  -\overline{\boldsymbol{p}}_2& -\overline{\boldsymbol{p}}_1 \end{pmatrix}~, &
U_E \cdot \cS &= i\sigma_3 \cS~.
\end{align}
 Taking the quotient $Y/G_E = X$ leads to the Enriques surface as the compactification geometry, but we see that it does not inherit any of the spinors.  Notice also that when extended to act on the spin bundle the action is no longer order $2$.  

On the other hand, we observe that the background on $Y$ has another $\Z_2$ symmetry, with generator 
\begin{align}
\Ut_E  = (g, e^{i\pi\bF_{\sleft}}\Pi) \in\SO(4)\times \cG_{\text{p}}~.
\end{align}
This action has a lift to $\Spindp(4)$ and acts on $\cS$ by
\begin{align}
\Ut_E\cdot \cS & = i \sigma _3 \cS i \sigma_2 = - \sigma_3 \cS \sigma_2~.
\end{align}
Now the action squares to identity, and there are invariant fermions that are solutions to
\begin{align}
\cS & = -\sigma_3 \cS \sigma_2~.
\end{align}
In this way we obtain an explicit geometric understanding of supersymmetry preservation on a non-spin background geometry.

\section{Further examples}
The conclusion of the previous section is that it is possible to compactify IIB string theory on the Enriques surface by using~$\Spindp$ fermions.  Moreover, it is possible to analyze the resulting theory both from supergravity and from the worldsheet, and the latter does not require any non-perturbative degrees of freedom.  In this section we will discuss a few other examples of these sorts of ``flat F-theory'' backgrounds.

\subsection{Toroidal generalizations}

The first idea one might have is to replace the K3 geometry with the simpler choice of $T^4$.  The analogue of the Enriques involution on periodic coordinates $x_i$, $x_i \sim x_i+1$ is then simply
\begin{align}
U_E \cdot (x_1,x_2,x_3,x_4) = \left(x_1+\ff{1}{2},-x_2,-x_3,x_4\right)~,
\end{align}
and we see that the space is of the form $X_3 \times S^1$, where $X_3$ is a compact flat Riemannian manifold with holonomy $\Z_2$.  Dropping the spectator circle, we see that there is a $5$-dimensional flat F-theory background, $(T^3 \times T^2)/\Z_2$, combining a Bieberbach action on the first factor with a holomorphic involution on the second factor.  This $7$-dimensional background was recently discussed in~\cite{Montero:2022vva}, where it is also pointed out that it has non-perturbative relatives with the $\Z_2$ action replaced by one of $\Z_3,\Z_4,\Z_6$.  While those quotients still do not have orientifold planes, the IIB axio-dilaton is now fixed to special values, and, moreover, the action on the worldsheet spectrum invariably mixes NS-NS with R-R fields.\footnote{In this case the non-trivial ``modular'' factors of the form $(c\tau+d)/(c\taub +d)$ that appear in the transformations of the fermions---see, for example, the discussion in section 3.1 of~\cite{Debray:2021vob}---must be included to obtain the desired invariant fermions.}  The latter feature makes it difficult to understand the world-sheet theory in the RNS formalism, and the former signals that a perturbative worldsheet treatment will be at best formal because the dilaton is fixed.

It is also possible to build similar lower-dimensional compactifications.  For instance, consider
$Z = ( (T^2)^4 \times S^1)/\Z_2$, where the action of the generator is
\begin{align}
g\cdot (z_0,z_1,z_2,z_3;x) = \left(-z_0,-z_1,-z_2,-z_3;x+\ff{1}{2}\right)~.
\end{align}
$Z$ has holonomy $\Z_2 \subset \Sp(1)\times\Sp(1)$, and it is a flat F-theory background with base the $7$-dimensional non-spin manifold $X =(T^6\times S^1)/\Z_2$ obtained by dropping the $z_0$ coordinate.   We then have the corresponding perturbative IIB orientifold of $T^7\times \R^{1,2}$ that preserves $8$ supercharges in three dimensions.

There is no need to restrict to a quotient by a single cyclic group.  Consider again $T^9$ and the action of $G = \Z_2\times\Z_2\times\Z_2$ with generators acting on the holomorphic coordinates as
\begin{align}
g_1 \cdot (z_0,z_1,z_2,z_3;x) & = \left( -z_0, -z_1,z_2+\ff{1}{2},z_3;x\right)~, \nonumber\\
g_2 \cdot (z_0,z_1,z_2,z_3;x) & = \left( -z_0, z_1,-z_2,z_3+\ff{1}{2};x\right)~, \nonumber\\
g_3 \cdot (z_0,z_1,z_2,z_3;x) & = \left( -z_0, z_1+\ff{1}{2},z_2,-z_3;x+\ff{1}{2}\right)~.
\end{align}
The action is free on $T^9$, as well as on the $T^7$ spanned by the coordinates $(z_1,z_2,z_3;x)$.  The only holomorphic form preserved is $dz_0\wedge dz_1 \wedge dz_2 \wedge dz_3$, so this is an F-theory background preserving $4$ supercharges in $3$ dimensions, and it has a perturbative orientifold description with a large radius limit as a IIB supergravity compactification on the non-spin manifold $T^7/G$.  

\subsection{Flat F-theory with curved target space} \label{ss:curvedexamples}
It is possible to generalize the discussion to curved geometries.  The first idea one might have is to replace the Enriques surface with a quotient of a higher-dimensional Calabi-Yau manifold with holonomy $\SU(n)$ by a freely-acting K\"ahler isometry that projects out the holomorphic $n$-form.  This is not possible for $n=3$: the Euler characteristic of the structure sheaf of the Calabi-Yau $3$-fold is zero, and the quotient necessarily preserves this vanishing.  It then follows that the holomorphic $3$-form will be preserved.\footnote{We thank James Gray for explaining this to us.}   When $n$ is even this restriction does not apply, and it may be possible to find a $4$-fold with an appropriate free $\Z_2$ action.  However, even for $n=2$ the constraint is non-trivial: for example it immediately shows that a freely-acting K\"ahler isometry of a K3 surface that projects out the holomorphic form must have order $2$.

Alternatively, we can allow for a non-free action on the Calabi-Yau $n$-fold but compensate this by a shift on an additional circle.  For example, consider IIB string theory compactified on $Y \times S^1$, where $Y$ is a K3 surface realized as a smooth quartic hypersurface in $\P^3$.  Denoting the projective coordinates by $[y_0:y_1:y_2:y_3]$, we take the hypersurface as the vanishing locus of 
\begin{align}
P = y_0^3 y_1 + y_1^4 + y_2^4 +y_3^4~.
\end{align}
$Y$ has a large group of K\"ahler isometries inherited from isometries of the ambient $\P^3$, including a $\Z_4$ subgroup generated by $g_4$, and $\Z_6$ subgroup generated by $g_6$, with
\begin{align}
g_4 \cdot [y_0 :y_1:y_2:y_3]  &= [y_0: y_1 :y_2 : \zeta_4 y_3]~, \nonumber\\
g_6\cdot [y_0:y_1:y_2:y_3] &= [-\zeta_3 y_0: -y_1: -y_2:y_3]~,
\end{align}
where $\zeta_n = e^{2\pi i/n}$ denotes the $n$-th primitive root of unity.  We recall---see, e.g.~\cite{Cox:2000vi}---that in this case the holomorphic (2,0) form has a residue presentation as
\begin{align}
\omega & =\Res 
\frac{1}{P} \sum_{j=0}^{3} (-1)^j y_j  dy_0 \wedge \cdots \wedge \widehat{dy_j}\wedge \cdots \wedge dy_{d-1}~,
\end{align}
which makes it easy to read off the action of the generators on $\omega$:
\begin{align}
g_4 \cdot \omega & = \zeta_4 \omega~, & g_6 \cdot \omega = \zeta_6 \omega~.
\end{align}
There are different ways in which we can use $Y$ for a string compactification.  For example, we could use the Borcea-Voisin construction and its generalizations to build Calabi-Yau $3$-folds of the form $Z = (Y\times T^2) /G$, with $G = \Z_n$ and $n\in\{2,3,4,6\}$.  This will typically require a resolution of singularities.\footnote{A review of the original construction, and further references can be found in~\cite{Cox:2000vi}, while more recent work on the $n>2$ case, as well as more general automorphisms of non-prime order includes~\cite{Cattaneo:2016bvg,Bell:2022nsp}.}   Alternatively, we define a modified action on $Y\times S^1$ with generators
\begin{align}
\gt_n = \left(g_n, x\mapsto x+ \ff{1}{n}\right)~.
\end{align}
The quotient by the $\Z_n$ action generated by $\gt_n$ is a smooth non-spin manifold $X$ with a K\"ahler contact structure.  But, now just as above we can construct a IIB orientifold of $Y\times S^1$, which will be a purely closed string theory and will preserve $8$ supercharges in $5$ dimensions and can also be viewed as a IIB compactification on $X$.  The corresponding F-theory background is the quotient of $ Z = (T^2 \times Y \times S^1)/\Z_n$, where the first factor is the elliptic fiber.

We can replace $Y$ by a Calabi-Yau $3$-fold and obtain an analogous construction of IIB compactification on $X$---a smooth non-spin $7$-dimensional manifold with a K\"ahler contact structure---which nevertheless preserves $4$ supercharges in three dimensions.

Going down further to ``compactification'' on a Calabi-Yau $4$-fold, we can obtain superficially similar results.  For instance there is a flat F-theory background $Z = (Y_1\times Y_2\times T^2)/G$, where each $Y_i$ is a K3 admitting an Enriques involution $\phi_i$.  Denoting the reflection on $T^2$ by $\rho$, we consider the $G=\Z_2\times\Z_2$ action on the total space generated by
\begin{align}
g_1 & = (\phi_1, \iden, \rho)~,&
g_2 & = (\iden, \phi_2,\rho)~.
\end{align}
The action is free and projects out the holomorphic $(1,0)$, $(2,0)$ and $(3,0)$ forms, but keeps the $(5,0)$ form, so that $Z$ is a smooth elliptically fibered Calabi-Yau $5$-fold.  The holonomy and fundamental group of the $5$-fold $Z$ can be worked out in a similar fashion to the Enriques Calabi-Yau reviewed in the appendix below.  The result is that $\pi_1(Z) = \Z^2 \rtimes G$, and the holonomy is $\Gamma = (\SU(2) \rtimes \Z_2)^2 \subset \SU(5)$.  Since $\Gamma$ is not contained in $\SU(4)$, we expect this to preserve (2,0) supersymmetry in two dimensions.  Further examples of this sort can be constructed by quotients of hyper-K\"ahler manifolds with holonomy $\Sp(2)$, such as the Enriques manifolds discussed in~\cite{Oguiso:2010emf,BOISSIERE:2011}.

We note that in all of the examples, flat or curved, the F-theory background has an elliptic fibration with section $z_0 = 0$.

\section{Unoriented IIA compactification}
The introduction of an additional circle direction allows us to also consider a T-dual type IIA picture.  While the considerations apply to all of the examples described above, in this section we discuss the basic case of  IIA compactified $Y \times S^1$, where $Y$ is a K3 manifold admitting the Enriques involution.   Taking an orientifold by worldsheet parity is now problematic, since the IIA GSO projection is chiral in the Ramond sector.  However, we also have T-duality of the additional circle with an action $T$, which acts on the internal right-moving NS fields of the circle as
\begin{align}
T : X^5_\sright & \mapsto -X^5_\sright~,&
T : \cX^5_{\sright} &\mapsto -\cX^5_{\sright}~,
\end{align}
while leaving the other NS fields invariant, and maps the IIB string on $Y\times S^1$, where $S^1$ has radius $R$, to the IIA string on $Y \times S^1$ with the T-dual radius $\alpha'/R$.  We extend this action to the remaining sectors so that $T^2 =e^{i\pi \bF_\sright}$, and
\begin{align}
T e^{i\pi \bF_{\sleft}} T^{-1} & = e^{i\pi \bF_{\sleft}}~,&
T U_E T^{-1} & = U_E~,&
T \Pi T^{-1} &= R_5 \Pi~, 
\end{align}
where $R_5$ is a reflection on the $5$-th circle.  This allows us to define a $\Z_2$ symmetry of the IIA string on $Y\times S^1$ with generator
\begin{align}
\Ut_{\text{IIA}} = T \Ut_E T^{-1} = T U_E e^{i\pi \bF_{\sleft}} \Pi T^{-1} = e^{i\pi \bF_{\sleft}} U_E R_5 \Pi~,
\end{align}
and gauging this symmetry leads to a consistent orientifold.  The action $U_ER_5 $ is free on $Y\times S^1$, so that once again there are no orientifold planes or open string degrees of freedom, but this time it projects out the volume form!  Consequently, the spacetime interpretation is now a compactification of IIA supergravity on the unoriented manifold $\Xt = (Y\times S^1)/\Z_2$, where the generator of $\Z_2$ is $(\phi,\rho)$, where $\phi$ is the Enriques involution on the K3 factor and $\rho$ is the reflection symmetry of the single circle.  As in our IIB discussion, the novelty is not so much in the loss of the orientation in the quotient---after all that is inherent in IIA orientifolds, as was pointed out long ago~\cite{Sen:1997kz}; instead, it is that the compactification has a completely string-perturbative description with a supergravity limit, and it is T-dual to the IIB compactification on the Enriques surface discussed above.\footnote{Similar arguments should apply to our other examples upon further compactification on a circle, but for the remainder of this section we will stick to $\Xt$.}

\subsection{A check of the massless spectrum}
The $5$-dimensional massless spectrum provides a simple check of the proposed T-dual pair.  This can be carried out directly in spacetime and in general requires the machinery of twisted cohomology, or, more precisely, cohomology with local coefficients---see, for example~\cite{Davis:1991alg} for a pedagogical introduction.  In our case the situation is much simpler, because we can work on the covering space and just project onto invariant states.

To see how this works we start with the massless spectrum of IIB supergravity compactified on the K3 surface $Y$ on the Enriques locus in the moduli space.  Returning to the cohomology of $Y$, we have as in~(\ref{eq:evenoddcohomology}) the non-vanishing groups
\begin{align}
H^0(Y,\R) & = \left(\R\right)_{\text{e}}~,&
H^2(Y,\R) & = \left(\R^{10}\right)_{\text{e}} \oplus\left(\R^{12}\right)_{\text{o}}~,&
H^4(Y,\R) & =  \left(\R\right)_{\text{e}}~,
\end{align}
while the action of the gauge symmetry $e^{i\pi \bF_{\sleft}}\Pi $ on the ten-dimensional bosonic fields is as follows:
\begin{align}
\text{field}	&&g	&&\phi	&&B		&&C_0	&&C_2	&&C_4^+		\nonumber\\ 
  e^{i\pi\bF_{\sleft}}\Pi~		&&+	&&+	&&-		&&+	&&-	&&+		\nonumber\\
\end{align}
To describe the massless bosonic spectrum it then suffices to reduce these fields while keeping the invariant forms and to supplement that counting with deformations of $g$ on the Enriques locus. We see that the reduction of $C_4^+$ yields $10$ tensor fields and a scalar, while $B$ and $C_2$ each lead to $12$ scalar fields.  Including the axio-dilaton and the $30$ real scalars from deformations of $g$ we obtain the six-dimensional massless bosonic content of the (1,0) supergravity multiplet, $9$ tensor multiplets and $12$ hypermultiplets.

We obtain the $5$-dimensional spectrum by reducing the theory further on $S^1$.  Upon reduction each tensor multiplet gives rise to a $5$-dimensional vector multiplet, each hypermultiplet descends to a $5$-dimensional hypermultiplet, and the (1,0) supergravity multiplet reduces to the minimal $5$-dimensional supergravity multiplet and a vector multiplet, so that the massless bosonic content consists of the metric, $11$ vectors, and $58$ real scalars.  The latter include the $31$ real scalars corresponding to deformations of the Enriques metric and the radius of the additional circle.

The same methods are easily applied to the massless spectrum of IIA string on the unoriented manifold $\Xt = (Y\times S^1)/\Z_2$.  Setting $\widetilde{Y} = Y\times S^1$, we have the cohomology groups
\begin{align}
H^0(\widetilde{Y},\R) & = \left(\R\right)_{\text{e}}~,&
H^1(\Yt,\R) & =  \left(\R\right)_{\text{o}}~,&
H^2(\Yt,\R) & = \left(\R^{10}\right)_{\text{e}} \oplus\left(\R^{12}\right)_{\text{o}}~,& \nonumber\\
H^5(\Yt,\R) & =  \left(\R\right)_{\text{o}}~, &
H^4(\Yt,\R) & =  \left(\R\right)_{\text{o}}~,&
H^3(\Yt,\R) & = \left(\R^{10}\right)_{\text{o}} \oplus\left(\R^{12}\right)_{\text{e}}~.&
\end{align}
The action of $e^{i\pi \bF_{\sleft}}R_5\Pi $ is
\begin{align}
\text{field}				&&g	&&\phi	&&B		&&C_1	&&C_3		\nonumber\\ 
  e^{i\pi\bF_{\sleft}}R_5\Pi~	&&+	&&+		&&-		&&-		&&+		\nonumber\\
\end{align}
To obtain the massless bosonic spectrum on $\Xt$ we again take the invariants obtained by reducing these fields, supplemented by $31$ deformations of the metric.  We see that $B_2$ and $C_3$ lead to $11$ vectors: one of these resides in the gravity multiplet, while the others are in $10$ $5$-dimensional vector multiplets.  The dilaton yields $1$ scalar, while the reduction of $B_2$, $C_3$ and $C_1$ leads to, respectively,  $12$, $13$, and $1$ scalars.  All in all, then, we find a metric $11$ vectors, and $58$ scalars, just as in the IIB description.

\subsection{Pinors and spacetime supersymmetry}
The worldsheet construction makes it clear that the IIA string yields a supersymmetric compactification on $\Xt$, but just as in the IIB discussion, we would like to have a spacetime interpretation of this, given that $\Xt$ is not orientable.   The point is that nevertheless $\Xt$ carries a $\text{Pin}^-$ structure, and this allows the existence of the requisite supersymmetric fermions.

As a first step we note that $\Xt$ is a circle bundle over the Enriques surface $X$: the projection $p_1 : \Xt \to X$ simply forgets the circle direction. The bundle has a section, where $i_1: X\to \Xt$ is given by $i_1(x) = (x,0)$, and of course $i_1 p_1 = \text{id}_{\Xt}$, $p_1 i_1 = \text{id}_X$.  It follows that
\begin{align}
T_{\Xt} = p_1^\ast (T_{X}) \oplus p_1^\ast (L)~,
\end{align}
where $L = \det T_X$.  We can use $p_1$ to pull back Stiefel-Whitney classes to conclude
\begin{align}
\label{eq:projectEnriques}
w_1 (T_{\Xt}) &= p_1^\ast w_1(L)~, &
w_2(T_{\Xt})  &= p_1^\ast\left( w_2(T_X) + w_1(T_X) \cup w_1(L)\right) = p_1^\ast w_2(T_X)~.
\end{align}
The last equality follows because $X$ is oriented, so that $w_1(T_X) = 0$.  Because $i_1^\ast p_1^\ast = \text{id}_{\Xt}$, we see that $w_1(L) \neq 0$ and $w_2(T_{\Xt}) \neq 0$.  Next, note that $\Xt$ itself is the base of a circle bundle $Z = (Y\times S^1\times S^1)/\Z_2$ --- the Enriques Calabi-Yau $3$-fold we encountered above.  Using the projection $p_2 : Z \to \Xt$ and section $i_2 :  \Xt \to Z$ defined in an analogous fashion, we then see
\begin{align}
0 & = w_1(T_Z) = p_2^\ast \left( w_1 (T_{\Xt}) + p_1^\ast w_1(L) \right)~, \nonumber\\
0 & = w_2(T_Z) = p_2^\ast \left( w_2 (T_{\Xt}) +w_1(T_{\Xt}) \cup p_1^\ast w_1(L)\right)~.
\end{align}
Using $i_2^\ast p_2^\ast = \text{id}_{\Xt}$ we conclude
\begin{align}
w_2(T_{\Xt}) & \neq 0~, &
w_2(T_{\Xt}) + w_1(T_{\Xt}) \cup w_1(T_{\Xt}) & = 0~.
\end{align}
But now we recall~\cite{Kirby:1991pin} that $w_2$ is the obstruction to the existence of a $\text{Pin}^+$ structure, while $w_2 + w_1 \cup w_1$ is the obstruction to the existence of a $\text{Pin}^-$ structure.  Thus, we see that $\Xt$ admits a $\text{Pin}^-$ structure but does not admit a $\text{Pin}^+$ structure.  In fact, with our simple quotient construction it is not hard to check explicitly that a reflection in the circle direction squares to $-\iden$ on the pinor over $\Xt$, which is the defining feature of a $\text{Pin}^-$ structure.

The $\text{Pin}^-$ structure on $\Xt$ can be thought of as a lift to the pinor bundle of the holonomy group of $\Xt$---$\SU(2)\rtimes \Z_2 $, which is contained in $\GO(5)$ but not in $\SO(5)$:  this bundle has non-zero covariantly constant sections, and this is responsible for the preservation of spacetime supersymmetry.

\subsection{M-theory lift}
What is the M-theory lift of IIA compactified on $\Xt$?  Since M-theory compactification only makes sense on a manifold with a $\text{Pin}^+$ structure, by the preceding discussion the lift cannot simply be $\Xt \times S^1$.  On the other hand, the duality with F-theory provides a clear answer:  the lift is simply M-theory compactified on the Enriques Calabi-Yau $Z$.  The IIA orientifold action can then be understood from the M-theory perspective as follows.  The appearance of $e^{i \pi \bF_{\sleft}}$ in the IIA reduction is explained by the observation~\cite{Diaconescu:2000wy} that the reflection symmetry of the M-theory circle corresponds to $e^{i\pi \bF_{\sleft}}$ in the IIA string.  On the other hand, the GSO projection in the IIA worldsheet description requires the reflection in the $10$-dimensional spacetime to be accompanied by worldsheet parity.

This construction can be compared with M-theory compactified on the Klein bottle~\cite{Dabholkar:1996pc}.  In that case we start with M-theory on a non-oriented manifold with a $\text{Pin}^+$ structure---as explained in, for example~\cite{Witten:2016cio,Tachikawa:2018njr,Freed:2019sco}, this is required for a consistent M-theory background---and its IIA reduction is the orbifold of $S^1$ that combines $e^{i\pi \bF_{\sleft}}$ and a half-circle shift.  Here, on the other hand, the M-theory background is spin, and its IIA reduction is a compactification on a non-orientable manifold with a $\text{Pin}^-$ structure and an orientable but non-spin IIB-dual.

\section{Outlook}
In this short note we pointed out a simple class of supersymmetric IIB string compactifications that share a number of interesting properties:  first, they have a perturbative closed string orientifold description; second, they are described as flat F-theory backgrounds, where the axio-dilaton profile remains constant over a K\"ahler base; finally, they have a supergravity description as compactifications of IIB supergravity on non-spin manifolds.  The latter feature depends on the non-trivial transformations of IIB fermion fields under the IIB duality group.  With the addition of a circle the backgrounds also have a IIA interpretation as compactification on unoriented manifolds.  

While these backgrounds are rather trivial from the point of view of generic F/M-theory compactifications, they deserve a closer look precisely because, while preserving various amounts of supersymmetry, they have a perturbative string-theoretic description and a geometric interpretation.  They thus offer a perfect framework for exploring aspects of F/M-theory away from the large radius limit, as well as string duality along the lines of~\cite{Vafa:1995gm}.  For instance, it may be possible to find  dual descriptions of some of our $5$-dimensional examples as heterotic compactification on compact flat manifolds that has been recently studied in~\cite{Cheng:2022nso}.  Perhaps they may help to explore new spacetime perspectives on string orbifold constructions such as that pursued in~\cite{Giaccari:2022xgs}.   

We envision several ways to explore and expand on these observations.  Perhaps the most immediate would be to explore the appropriate generalization of spacetime T-duality that would account for the non-trivial transformation from a compactification on a manifold with non-vanishing $w_2$ Stiefel-Whitney class to a T-dual compactification on a non-orientable manifold.  While the appearance of non-spin manifolds has been discussed in the context of topological T-duality, for example in~\cite{Bouwknegt:2003vb}, the resolution of the attending supersymmetry puzzle involved a non-trivial flux and a corresponding $\text{Spin}_{\text{c}}$ structure.  It would be very useful to rephrase our T-dual pairs in a purely spacetime language that generalizes the results of~\cite{Bouwknegt:2003vb}.

It will also be good to have a better understanding of the F- and M-theory lifts of these string backgrounds.  In each case we see that the lift gives a ``geometrization'' of the action of a spacetime gauge symmetry, such as $\Pi e^{i\pi\bF_{\sleft}}$ in the string compactification, resulting in a locally constant but globally non-trivial fibration of the M-theory circle or the F-theory torus. What are the general rules?

In addition it will be interesting to look for string dual descriptions.  Are there manifestations of the ``non-spin'' nature of these compactifications in the dual description?  It would also be interesting to explore the constructions more systematically.  For example, one could use the classification of compact flat K\"ahler manifolds~\cite{Dekimpe:2009kfm} to identify suitable non-spin manifolds and study in detail the resulting orientifolds.  

More generally, the global structures studied here can also arise in less trivial examples that also involve fluxes and might lack an explicit worldsheet description.  To handle such situations it would be useful to develop a theory akin to that of G-structures that would also incorporate the global effects.  Finally, it should also be possible to find other routes to non-spin compactifications.  For instance, by compactifying the heterotic string on $T^2$ one also obtains a duality group action, and further compactification could make use of this.  Studying such compactifications, their geometries, non-geometric limits, and dual descriptions should provide valuable markers in the string landscape.

\appendix
\section{$G$-structures and the Enriques Calabi-Yau manifold} \label{app:GstrFHSV}
In this short appendix we make some remarks on the topology of the Enriques Calabi-Yau $Z = (Y\times T^2)/\Z_2$.  A detailed study of the full homology was carried out in~\cite{Aspinwall:1995mh}.  We will not need all of those details and just note two basic facts: the fundamental group is given by $\pi_1(Z) = (\Z)^2 \rtimes \Z_2$, while its holonomy group is $\Gamma = \SU(2)\rtimes\Z_2 \subset \SU(3)$.    

To explain the first result we can work on the covering space $\widetilde{Z} = Y \times \R^2$.   To obtain $Z$ we quotient by a group with generators $h[\vec{m}]$, $\vec{m} \in \Z^2\subset \R^2$, and $g$, which act on a point $z = (p,\vec{x}) \in \widetilde{Z}$ as
\begin{align}
h[\vec{m}] (p,\vec{x}) &= (p,\vec{x}+\vec{m})~, &
g (p,\vec{x}) & = ( \phi(p), -\vec{x})~,
\end{align}
where $\phi: Y \to Y$ is the Enriques involution.  From this form of the generators we easily see that $h[\vec{m}]$ generate a normal subgroup $\Z^2$, while $g$ generates a subgroup $\Z_2$, and the group product is the advertised semi-direct product:
\begin{align}
h[\vec{m}] g  h[\vec{n}] g^k = h[\vec{m}-\vec{n}] g^{k+1}~.
\end{align}

Moving on to the holonomy, the first factor is associated to $\Gamma_0$, the holonomy group generated by parallel transport around null-homotopic loops, while the second is associated to the non-trivial homotopy class corresponding to the generator  $g\in \pi_1(Z)$.\footnote{As reviewed in~\cite{Joyce:2000cm}, $\Gamma_0\subset \Gamma$ is a normal subgroup, and in general there is a surjective homomorphism $\pi_1(\cY) \to \Gamma/\Gamma_0$.  In our example the holonomy for any loop associated to the $\Z^2$ factor is contained in $\SU(2)$, and so is trivial in $\Gamma/\Gamma_0$.  
}  
We can understand the appearance of the semi-direct product by considering the holonomy action on $Z$ in a little more detail. First, fixing a base-point and a holomorphic frame for the cotangent bundle $\{\boldsymbol{p}_1,\boldsymbol{p}_2,\boldsymbol{p}_3\}$, where the first two basis vectors are as in~(\ref{eq:holoframe}) while $\boldsymbol{p}_3$ is the holomorphic frame associated to the torus fiber, the holonomy generated by the null-homotopic loops is $\Gamma_0 = \SU(2)$, and its elements are of the form
\begin{align}
A[M] & = \begin{pmatrix} M & 0 \\ 0 & 1 \end{pmatrix}~,
\end{align}
where $M$ is an $\SU(2)$ matrix in the fundamental representation.  On the other hand, a representative for the holonomy along the loop corresponding to $g$ is (up to conjugation by an $A[M]$) given by
\begin{align}
B & = \begin{pmatrix} -\sigma_3 & 0 \\ 0 & -1\end{pmatrix}~.
\end{align}
Again, the explicit product of elements of the form $A[M] B$ shows the semi-direct product structure.  Note that we obtain a very similar discussion if we drop the $T^2$ fiber all together:  we will then find that the Enriques surface has fundamental group $\Z_2$ and holonomy $\SU(2)\rtimes\Z_2 \subset \GU(2) \subset \SO(4)$.  

It is easy to extend this analysis to the $5$-fold example at the end of section~\ref{ss:curvedexamples}, where now the holonomy group is generated by
\begin{align}
A_1[M] & = \begin{pmatrix} M & 0 &0 \\ 0 & \iden_2 & 0 \\ 0 & 0 & 1 \end{pmatrix}~,&
A_2[M] & = \begin{pmatrix} \iden_2 & 0 &0 \\ 0 & M & 0 \\ 0 & 0 & 1 \end{pmatrix}~,\nonumber\\[2mm]
B_1 & = \begin{pmatrix} -\sigma_3 & 0 &0 \\ 0 & \iden_2 & 0 \\ 0 & 0 & -1\end{pmatrix}~, &
B_2 & = \begin{pmatrix} \iden_2 & 0 &0 \\ 0 & -\sigma_3 & 0 \\ 0 & 0 & -1\end{pmatrix}~.
\end{align}
Since $A_1[M]$ and $B_1$ each commute with $A_2[M]$ and $B_2$, the holonomy is $(\SU(2)\rtimes\Z_2)^2$.

Finally, we make one remark concerning the relationship between the preservation of multiple $G$-structures and supersymmetry.   The quotient preserves two $\SU(3)$ \textit{structures} on $Z$, one with $3$-form $\Omega = \omega \wedge dz$, and another with $3$-form $\Omega' = \omega \wedge d\zb$.  This is a feature often seen in supergravity solutions with enhanced supersymmetry:  different spinors stabilize different $\SU(3)$ structures, and their intersection can lead to a smaller structure group.  In this case the two $\SU(3)$ structures indeed intersect in a smaller group  $\SU(2)\rtimes \Z_2 \subset\SU(3)$, but it is just large enough to avoid enhanced supersymmetry.

\bibliographystyle{./utphys}
\bibliography{./newref}

\end{document}

%% file: basedefs2.tex
\DeclareFontFamily{U}{rsf}{}
\DeclareFontShape{U}{rsf}{m}{n}{
  <5> <6> rsfs5 <7> <8> <9> rsfs7 <10-> rsfs10}{}
\DeclareMathAlphabet\Scr{U}{rsf}{m}{n}

\def\CO#1#2{{[#1,#2]}}
\def\AC#1#2{{\{#1,#2\}}}

\def\iden{{\mathbbm 1}}

\def\rep#1{{{\boldsymbol{#1}}}}

\def\C{{\mathbb C}}

\def\P{{\mathbb P}}
\def\R{{\mathbb R}}
\def\Z{{\mathbb Z}}

\def\Gr{\operatorname{Gr}}

\def\Res{\operatorname{Res}}

\def\dVol{\operatorname{dVol}}

\def\SO{\operatorname{SO}}
\def\SL{\operatorname{SL}}
\def\GL{\operatorname{GL}}

\def\GO{\operatorname{O{}}}
\def\SU{\operatorname{SU}}
\def\GU{\operatorname{U{}}}
\def\Sp{\operatorname{Sp}}
\def\Spin{\operatorname{Spin}}

\def\so{\operatorname{\mathfrak{so}}}

\def\spin{\operatorname{\mathfrak{spin}}}
\def\su{\operatorname{\mathfrak{su}}}

\def\p{\partial}
\def\pb{\bar{\partial}}

\def\ff#1#2{{\textstyle\frac{#1}{#2}}}

\def\cA{{\cal A}}

\def\cC{{\cal C}}

\def\cF{{\cal F}}
\def\cG{{\cal G}}

\def\cJ{{\cal J}}

\def\cM{{\cal M}}

\def\cO{{\cal O}}
\def\cP{{\cal P}}
\def\cQ{{\cal Q}}
\def\cR{{\cal R}}
\def\cS{{\cal S}}

\def\cU{{\cal U}}

\def\cX{{\cal X}}
\def\cY{{\cal Y}}

\def\ep{{\epsilon}}

\newcommand\taub{\overline{\tau}}

\newcommand\omegab{\overline{\omega}}

\newcommand\sigmat{\widetilde{\sigma}}

\newcommand\Psib{\overline{\Psi}}

\newcommand\ib{\overline{\imath}}
\newcommand\jb{\overline{\jmath}}
\newcommand\kb{\overline{k}}

\newcommand\mb{\overline{m}}

\newcommand\zb{\overline{z}}

\newcommand\gt{\widetilde{g}}

\newcommand\Jh{\widehat{J}}

\newcommand\Lh{\widehat{L}}

\newcommand\Zb{\overline{Z}}

\newcommand\Ut{\widetilde{U}}

\newcommand\Xt{\widetilde{X}}
\newcommand\Yt{\widetilde{Y}}